\documentclass[preprint,12pt]{elsarticle}
\journal{journal} \RequirePackage{graphicx}

\begin{document}

\begin{frontmatter}

\title{Compact stars with specific mass function}

\author{S.K. Maurya}
\address{Department of Mathematical and Physical Sciences,
College of Arts and Science, University of Nizwa, Nizwa, Sultanate
of Oman\\sunil@unizwa.edu.om}

\author{Y. K. Gupta}
\address{Department of Mathematics, Raj Kumar Goel Institute of Technology, Ghaziabad,U.P.(India)\\
kumar001947@gmail.com}

\author{Farook Rahaman}
\address{Department of Mathematics, Jadavpur University, Kolkata 700032, West Bengal, India\\ rahaman@iucaa.ernet.in}

\author{Monsur Rahaman}
\address{Department of Mathematics, Jadavpur University, Kolkata 700032, West Bengal, India\\ mansur.rahman90@gmail.com}

\author{Ayan Banerjee}
\address{Department of Mathematics, Jadavpur University, Kolkata 700032, West Bengal, India\\ ayan\_7575@yahoo.co.in}

\date{Received: date / Accepted: date}

\maketitle

\hrule

{\bf H~I~G~H~L~I~G~H~T~S}

$*$ We have considered five dimensional flat spacetimes with coordinates $z^1$, $z^2$, $z^3$, $z^4$ and $z^5$.

$*$ We embedded five dimensional flat spacetimes into a four dimensional spacetimes by using the transformation.

$*$ This transformation gives a relation between metric potentials. After specifying this relation we construct the mass function.

$*$ We determined metric potential using mass function and generate physically plausible compact star with above specific mass function for anisotropic matter distribution.

\hrule

\begin{abstract}
\textbf{Aims:} In the present work we search for a new model of compact star within embedding class one spacetime i.e., four dimensional spacetime embedded in five dimensional Pseudo Euclidean space.

\noindent \textbf{Methods:} In particular we propose a new mass function to obtain an exact analytic solutions of the Einstein field equations. For this specific mass function, obtained solutions are well-behaved at the centre of the star, satisfy all energy conditions and the mass-radius relation fall within the limit proposed by Buchdahl \cite{Buchdahl}.

\noindent \textbf{Results:} The static equilibrium condition has been maintained under different forces. We have discussed the solutions in detail and compare with a set of astrophysical objects like 4U1608-52,  PSR J1903+327, PSR J1614-2230 and X-ray pulsar Vela X-1 is also explored.
\end{abstract}


\end{frontmatter}

\section{Introduction}

Stars are formed in gas and dust clouds in space where the matter density is slightly higher than in its surroundings.  In astronomy,  studying the structure and properties of compact star has attracted much attention to the researchers which is used to refer collectively to white dwarfs, neutron stars, and black holes.  It was realized long ago by Schwarzschild who obtained the exact solution of Einsteins field equation for the interior of a compact object.
The most probable compact stars are observed as pulsars, spinning stars with strong magnetic fields and build entirely of quark matter (QM). Finally, in  2006 Rosat surveys \cite{Agueros}, isolated compact stars by their X-ray emission that means the gravitational energy that gets released is radiated in X-rays.  The possible existence of pulsars were discovered by Bell-Burnell and Hewish \cite{Hewish,Pilkington}, that emits a beam of electromagnetic radiation which is actually continuous but beamed. Thus the discovery of pulsars and x-ray sources inspire physicists to develop theoretical models of compact stars like neutron stars (NS) and quark stars (QS). Though it is still unknown  the composition of particles  and nature of interactions,  in spite of several models are proposed time to time to describe compact stars.

In general, to analysis the properties and structure of compact stars, one needs to understand the equation of state (EOS) which relates the pressure as a thermodynamical function of the density. It is argued that EOS describing quark matter having a free degenerate Fermi gas of u, d, s quarks with equal masses, but  it is incompetent to expound the existence of massive neutron stars. From a theoretical viewpoint, the stability of highly-compact star with masses  around $2 M_{\odot}$ can be well explained with stiff EOS describing normal nuclear matter at high densities.
The mass of a white dwarf is usually approximately 0.5 - 0.6 $M_{\odot}$ having densities of order $\rho ~\sim$
$10^{10}$ kg/$m^3$, whereas neutron stars have $10^{18}$ kg/$m^3$. Thus, in order to estimated maximum size versus mass criterion of several compact objects such as PSR J1614-2230, PSR J1903+327 ,Vela X-1, 4U 1538-52 etc., some low-mass  X-ray  binaries, e.g., 4U  1636-536 and X-ray sources 4U 1728-34, PSR 0943+10, which
are not compatible with the standard neutron star models. From recent observational prediction there are more massive stars with matter densities of the order $10^{15}$ gr/cc, or higher, exceeding the nuclear matter density.
Addressing, such high matter densities star, the most possible hints that pressure to be anisotropic in general,
i.e., in the interior of such stars the radial pressure and tangential pressure are different. The history of anisotropic relativistic matter in general relativity goes back to the work done by several authors \cite{Bowers,Herrera11,Sharma1}
The structure of anisotropic compact stars and their phenomenological properties have been discussed in more detailed in
\cite{Maurya1,Maurya2,Maurya3,Abdul1,Farook1,Ayan1}. However, the existence of charged anisotropic compact star configurations
are considered in \cite{Islam,Piyali1,Newton1,Maurya11}, and in the same scenario extensive studies have been carried out
in the presence of different modified gravities \cite{Das1,Das2,Momeni1,Momeni2}. On the other hand Malaver \cite{Malaver11,Malaver22} have obtained static  solutions through gravitational potential  $Z(x)$ with electrical field for an isotropic and anisotropic matter distribution respectively.

According to recent studies on compact object can be analysed by
considering (i) an equation of state (ii) simple spatial geometry characterised by spheroidicity parameter,
and (iii) curvature parameter (R), however only a few solutions satisfy the required general physical criterions.
Therefore, exact solutions of the field equations that describe the interior of general relativistic stars
is still an active area of research. In other words, our main motivation in this paper is modeling
anisotropic stellar configuration satisfying Karmarkars condition i.e., spherically symmetric metric of embedding
class one. We refer our reader to \cite{Maurya4,Bhar1,Maurya5,Singh1,Maurya6,Gupta1} for a discussion on the embedding class I spacetime. The
main feature of embedding class I metric is the metric functions $\lambda$ and $\nu$ are correlated to each
other. We not only restricted ourself in modelling the diagram but also the entire analysis has been performed with a set of astrophysical objects in connection to direct comparison of some strange/compact star candidates
like X-ray burster 4U1608-52, X-ray sources PSR J1903+327, PSR J1614-2230 and X-ray pulsar Vela X-1.

The paper is organized as follow : in Sec. \textbf{2} the Einsteins field equations governing
the static spherically symmetric has been laid down. In Sec \textbf{3} the four-dimensional spacetime embedded in five-dimensional pseudo Euclidean space known as class I is determined.  We choose a physically meaningful
form of the mass function to obtain an exact analytic solutions of the Einstein field equations
in Sec. \textbf{4}.  In Sec \textbf{5}  we match our interior solution to the exterior vacuum solution at the
junction interference with radius r= R. In Sec \textbf{6} we determine the gravitational redshift of our model and
other physical properties describing in detail, where as in   Sec. \textbf{7} the physical properties
with the stability conditions have been studied. Finally, in Sec. \textbf{8}
we focuses on the some specific comments regarding the results obtained in the study.

\section{Einstein field equation for anisotropic fluid distribution:}

We begin with the space-time describing the interior of a spherically symmetric
star with zero angular momentum in the form:
\begin{equation}
ds^{2}=e^{\nu(r)}dt^{2}-e^{\lambda(r)}dr^{2}-r^{2}\left(d\theta^{2}+\sin^{2}\theta d\phi^{2} \right)\label{metric},
\end{equation}

where $\lambda$ and $\nu$ are the functions of the radial coordinate $r$,  are yet to
be determined.

We suppose that the energy-momentum tensor of the material composition
filling the interior of the compact star is a perfect fluid type and
the  Einstein's equations for locally anisotropic fluid are (we set G = c = 1):
\begin{equation}
 8\pi\,\rho=\frac{1}{r^2}-e^{-\lambda}\,\left[\frac{1}{r^2}-\frac{\lambda'}{r}\right],  \label{rho}
\end{equation}

\begin{equation}
 8\pi\,p_r=-\frac{1}{r^2}+e^{-\lambda}\,\left[\frac{1}{r^2}+\frac{\nu'}{r}\right], \label{radialp}
\end{equation}

\begin{equation}
 8\pi\,p_t=\frac{e^{-\lambda}}{4}\,\left[2\nu''+\nu'^2-\lambda'\,\nu'+\frac{2(\nu'-\lambda')}{r}\right]  \label{tangentialp}
\end{equation}
where `prime' denotes the derivative with respect to the radial co-ordinate, $r$.
Here, the energy density is denoted by $\rho(r)$ and  $p_r(r)$, $p_t(r)$ denotes fluid pressures along
the radial and transverse directions, respectively. Obtaining an exact solution we have four equations, namely, the
field equations (2)-(4), with five unknown functions of r, i.e., $\nu$, $\lambda$, $\rho(r)$, $p_r(r)$ and $p_t(r)$.
Obtaining an explicit solutions to the Einstein field equations we assumed a specific form
of mass function by imposing the Karmarkar condition.

\section{Class one condition for spherical symmetric metric}

At this stage we will determine the class one condition of the metric above (\ref{metric}),
 we suppose the 5-dimensional flat metric as:

\begin{equation}	ds^{2}=-\left(dz^1\right)^2-\left(dz^2\right)^2-\left(dz^3\right)^2-\left(dz^4\right)^2-\left(dz^5\right)^2,\label{e1}
\end{equation}

where we transforms corresponding elements of the Cartesian coordinate to spherical coordinates
$ z^1=r\,sin\theta\,cos\phi$, ~~ $z^2=r\,sin\theta\,sin\phi$, ~~$z^3=r\,cos\theta$, ~~$z^4=\sqrt{K}\,e^{\frac{\nu}{2}}\,cosh{\frac{t}{\sqrt{K}}}$ ~~and~$z^5=\sqrt{K}\,e^{\frac{\nu}{2}}\,sinh{\frac{t}{\sqrt{K}}}$.

On inserting the components of $ z^1,z^2, z^3, z^4 $ and $z^5$ into the metric (\ref{e1}), we get:
\begin{equation}
ds^{2}=-\left(\,1+\frac{K\,e^{\nu}}{4}\,{\nu'}^2\,\right)\,dr^{2}-r^{2}\left(d\theta^{2}+\sin^{2}\theta d\phi^{2} \right)+e^{\nu(r)}dt^{2},\label{e2}
\end{equation}

The equations above implies that the class of metric is 1 because we have
embedded 4-dimensional space time into 5-dimensional flat space time.

  The metric (\ref{e2}) may represent spacetime of embedding class one, if it holds the following
relations (see Ref. Maurya et al.\cite{Maurya4} for more details discussion)
\begin{equation}
e^{\lambda}=\left(\,1+\frac{K\,e^{\nu}}{4}\,{\nu'}^2\,\right),\label{eq4}
\end{equation}
where K is a constant of integration and
can easily establish a relationship between $e^{\lambda}$ and $e^{\nu}$.
After solving Eq. (\ref{eq4}), we get the value of $\nu$ as:
\begin{equation}
	\nu=2\,ln\left[A+\frac{1}{\sqrt{K}}\int{\sqrt{(e^{\lambda(r)}-1)}dr}\right], \label{nu1}
\end{equation}
where $A$ is arbitrary constant of integration.

\section{Anisotropic solution for compact star}

In order to study the static spherically symmetric configurations with anisotropic matter distribution
we suppose that $e^{-\lambda} = 1 - 2m/r$, which related to the quasilocal mass function m(r) by

\begin{equation}
e^{-\lambda} = 1 - \frac{2m}{r} = (1+K\,F'^2)^{-1}  \label{mass1}
\end{equation}

where $F=e^{\nu/2}$. Since mass $m(r)$ should satisfy the
conditions $m(0)=0$ and $m'(r)> 0$. From the Eq. (9) we have

 \begin{equation}
m'(r) = \frac{K\,[K\,F'^2\,(1+K\,F'^2)+ 2\,r\,K\,F'\,F''] }{((1+K\,F'^2)^2},
\end{equation}

which implies that
\begin{equation}
~~~ F^{\prime} F^{\prime\prime} > 0, ~ ~ ~ then ~~~ m'(r) > 0      \label{F11}
\end{equation}

If the condition Eq. (11) is satisfied then $m(r)$ will be positive and increasing function of $r$. For this purpose, we suppose the mass function is of the form

\begin{equation}
m(r) = \frac{ar^3\,tanh^2(br^2+c)}{2(1+ar^2\,tanh^2(br^2+c))},
\end{equation}
where $a $, $b $ and $c $ are non zero constant and $m(0)=0$. In general, to avoid the formation of the event horizon
within the spherical body and regular at stellar interior.

It is notable that the obtained relation leads to the following relationship
\begin{equation}
	e^{\lambda}=[1+ar^2\,tanh^2(br^2+c)].
\end{equation}

By plugging the value of $e^\lambda$ into Eq. (\ref{nu1}), we get

\begin{equation}
	F=e^{\nu/2}=\left[A+B\, \ln cosh(br^2+c)\right], \label{F12}
\end{equation}

with $B=\frac{\sqrt{a}\,}{2\,b\,\sqrt{K}}$. Now using Eq.(\ref{F12}), we have

\begin{equation}
	F^{\prime}= \frac{\sqrt{a}\,r}{\sqrt{K}}\, \tanh(b\,r^2+c)~~and~~ F^{\prime\prime}=\sqrt{a} [\tanh(br^2+c)+2\,br^2\,sech^2(br^2+c)].  \label{F122}
\end{equation}

since $F^{\prime}>0$ and  $F^{\prime\prime} > 0$ then $F'\,F^{\prime\prime} > 0$
which implies that mass $m(r)$ is increasing throughout the star.

 Now, we are going to find the components for stress-energy tensor which are extremely
complicated, but assuming a more simplified form of mass function these can be given in the form and
 have the following nonzero components

\begin{equation}
8\,\pi\,p_r=-\frac{\tanh \psi\left[-4\,b\,B+a\left(A+B \ln\cosh\psi\right)\,\tanh\psi\right]}
{\left[A+B\,\ln\cosh\psi\right]\,[1+a\,r^2\,\tanh^2\psi]^{2}},
\end{equation}
\begin{equation}
  8\,\pi\,p_t=\frac{2\,b\,r^2\,sech^2\psi\,\left[2\,b\,B-a\,(A+B\,\ln\cosh\psi)\,\tanh\psi\right]+\tanh\psi\,\,P_1(\psi)}
  {\left[A+B\,\ln\cosh\psi\right]\,\left[1+a\,r^2\,\tanh^2\psi\right]^{2}}.
\end{equation}
\begin{equation}
 8\,\pi\,\rho=\frac{a\,\tanh\psi\,\left[4\,b\,r^2\,sech^2\psi+3\,\tanh\psi+a\,r^2\,\tanh^3\psi\right]}
{\left[1+a\,r^2\,\tanh^2\psi\right]^{2}},
\end{equation}
An important quantity is to consider the pressure anisotropy of the fluid comprising the
dense star which leads to the following form
\begin{equation}
 8\,\pi\,\Delta=\frac{r^2\,\left[-2\,b\,B\,\cosh\psi+a\,(A+B\,\ln\cosh\psi)\,\sinh\psi\right]\,\left[a\,\sinh^3\psi-2\,b\,\cosh\psi\right]}
 {\cosh^4\psi\,\left[A+B\,\ln\cosh\psi\right]\,\left[1+a\,r^2\,\tanh^2\psi\right]^{2}},
\end{equation}
where $\Delta/r$ represents a force arising due to the anisotropic nature of the stellar model.
The anisotropy will be repulsive or directed outwards if $p_t > p_r$, and attractive or directed inward
when $p_t < p_r$. It is considerable that one can obtain an isotropic pressure corresponds to $\Delta=0$.

 However, it is important to emphasize the results more closely. That's why we
extend the mathematical forms for the energy momentum tensor, which are

\begin{equation}
\frac{dp_r}{dr}=2\,r\times\frac{f_{1}(\psi)\,f_{2}(\psi)+\tanh^2\psi\,[\,f_{3}(\psi)+a\,f_{4}(\psi)\,]}
{8\,\pi\,\left[A+B\,\ln\cosh\psi\right]^{2}\,\left[1+a\,r^2\,\tanh^2\psi\right]^{2}},
\end{equation}
\begin{equation}
\frac{dp_t}{dr}=r\times\frac{f_{5}(\psi) sech^6\psi [f_{6}(\psi)+f_{7}(\psi)+f_{8}(\psi)]+[f_{9}(\psi)+f_{12}(\psi)] [f_{10}(\psi)+f_{11}(\psi)]}
{4\,\pi\,\left[A+B\,\ln\cosh\psi\right]^{2}\,\left[1+a\,r^2\,\tanh^2\psi\right]^{3}},
\end{equation}
\begin{equation}
\frac{d\rho}{dr}=-2r\times\frac{f_{13}(\psi)+f_{14}(\psi)}{8\,\pi\,\left[1+a\,r^2\,\tanh^2\psi\right]^{3}},
\end{equation}
where, $P_1(\psi)=\left[4\,b\,B-a\,(A+B \ln\cosh\psi)\,\tanh\psi + 2\,a\,b\,B\,r^2\,\tanh^2\psi\right]$, \\

 $f_{1}(\psi)=-b\,\left[\,A+B\,\ln\cosh\psi\,\right]\,sech^4\psi$,\\

$f_{2}(\psi)=\left[-2bB(1+ar^2)-2bB(1-a r^2)\cosh2\psi+a(A+B\ln\cosh\psi)\sinh2\psi \right]$,\\

$f_{3}(\psi)=\left[\,-4\,b^{2}\,B^{2}-4\,a\,b\,B\,(A+B\,\ln\cosh\psi)\,\tanh\psi \,\right]$,\\

$f_{4}(\psi)=\left[a\,A^2-4 b^2\,B^2 r^2+2 a A B \ln\cosh\psi+a B^2 (\ln\cosh\psi)^{2} \right] \tanh^2\psi $,\\

$f_{5}(\psi)=\frac{b}{16\,\cosh^6\psi\,}\,\left[1-a\,r^2+ (1+a\,r^2)\,\cosh2\psi\right]\,(A+B\,\ln\cosh\psi)$,\\

$f_{6}(\psi)=\left[32\,b\,B-32\,a\,A\,b\,r^2-16\,a\,b\,B\,r^2(1+2\,\ln\cosh\psi)\right]$,\\

$f_7(\psi)= 16\,b\,\cosh2\psi\,\left[2\,B+a\,A\,r^2+a\,B\,r^2+a\,B\,r^2\,\ln\cosh\psi \,\right]$,\\

$f_8(\psi)=-2\left[8\,a\,A +a\,B +16\,b^2\,B\,r^2+8\,a\,B\,\ln\cosh\psi\right]\,\sinh2\psi+a\,B\,\sinh4\psi$,\\

$f_9(\psi)=-2\,a\,(A+B\,\ln\cosh\psi)\,\tanh\psi\, \left[2\,b\,r^2\,sech^2\psi+\tanh\psi\right]$,\\

$f_{10}(\psi)=\left[2\,b\,r^2\,sech^2\psi\,(2\,b\,B-a(A+B\,\ln\cosh\psi))\,\right]\,\tanh\psi+\tanh\psi $,\\

$f_{11}(\psi)=\tan\psi\,\left[\,4\,b\,B-a\,(A+B\,\ln\cosh\psi)\,\tanh\psi+2\,a\,b\,B\,r^2\,\tanh^2\psi\, \right]$,\\

$f_{12}(\psi)=\left[-b\,B\,\tanh\psi\,(1+a\,r^2\,\tanh^2\psi)\,\right]$,\\

$f_{13}(\psi)=a\,\left[a \tanh^4\psi\,(5+a\,r^2\,\tanh^2\psi)\right] + 4\,b^2\,r^2\,sech^4\psi\,(-1+3\,a\,r^2\,\tanh^2\psi)$,\\

$f_{14}(\psi)=2bsech^2\psi\tanh\psi\,\left[-5+4 b r^2 \tanh\psi+3 a r^2 \tanh^2\psi+4 a b r^4\,\tanh^3\psi \right] $,\\

$\left(br^2+c\right)= \psi$.\\

Finally, we move on to describe the results obtained from our calculations, which are illustrated
in Figs.  (1-3), describing the metric functions, energy density, radial and transverse pressures
and measure of anisotropic within the given radius. The present work just argues for obtaining
models for compact objects,  thereafter a qualitative analysis of the
physical aspects is considered in more detail.

\section{Matching condition}
Before moving on to the study of physical properties of compact star,
we match the interior solution  at the boundary r = R to the asymptotically flat vacuum
exterior Schwarzschild solution. This is the condition we impose the
radial pressure at boundary i.e., $p_r(R)=0$, gives
\begin{equation}
 \frac{A}{B}=\frac{\coth\Psi\,\left[4\,b-a\,\ln\cosh\Psi\,\tanh\Psi\right]}{a}.
 \end{equation}
More explicitly for this purpose, we set the condition  $1-\frac{2M}{R} =e^{\nu(R)}=e^{-\lambda(R)}$,
which yields
 \begin{equation}
 B=\frac{1}{\sqrt{1+a\,R^2\,\tanh^2\Psi}\,\left[\,\frac{A}{B}+\ln\cosh\Psi\,\right]}.
 \end{equation}
Also, from Eq. (12) implies that the total mass of star can now be
written as
  \begin{equation}
 M=\frac{a\,R^3\,\tanh^2\Psi}{2\,[1+a\,R^2\,\tanh^2\Psi]},
 \end{equation}
where, $\Psi=b\,R^2+c$.

\section{Physical analysis of our solution}

In this work our main emphasis is on the analysis of the
physical properties and characteristics of central density, radial and transverse pressures
inside the compact objects which leads to

\begin{equation}
\rho_c  = \frac{3\,a\,tanh^2\left(c\right)}{8\pi}~and ~ p_{rc} =  p_{tc}  = -\frac{\left[-4 b B+a\left(A+B \ln\cosh\left(c\right)\right) \tanh\left(c\right)\right]}
{8\pi \,\coth \left(c\right)\, \left[A+B\,\ln\cosh\left(c\right)\right]},
\end{equation}
Moreover, Zeldovich condition demands that the ration of central density and central pressure
must be $\le 1$, which reads
\begin{eqnarray}
\frac{4\,b\,B-a\left(A+B \ln\cosh\left(c\right)\right)\,\tanh\left(c\right)}{3\,a\tanh\left(c\right)\left[A+B\,\ln\cosh\left(c\right)\right]} \le 1, \label{cons2}
\end{eqnarray}
Using Eqs. (23) and (24) we get an inequality which reduces to
\begin{eqnarray}
-\frac{A}{B}\le\frac{a\,\tanh(c)\,\ln\cosh(c)-b}{a\,\tanh(c)} . \label{cons3}
\end{eqnarray}
Now we are in the position to calculate the compactification factor for our model which
related to the quasilocal mass function m(r) by
\begin{eqnarray}
u(r) & =& {m(r) \over r}= \frac{ar^2\,\tanh^2(br^2+c)}{2\,[1+ar^2\,\tanh^2\left(br^2+c\right)]},\label{co}
\end{eqnarray}
also, the redshift is defined by the relation
\begin{eqnarray}
z(r) &=& e^{-\nu/2}-1 = \frac{1-\left[A+B\,\ln\cosh\left(b\,r^2+c\right)\right]}{\left[A+B\,\ln\cosh\left(b\,r^2+c\right)\right]},
\end{eqnarray}
whence we obtain
\begin{eqnarray}
z_s& = \sqrt{1+a\,R^2\,\tanh^2\left(b\,R^2+c\right)}-1. \label{su}
\end{eqnarray}

Having derived the equations that describe strange compact stars, we now proceed to observe
them more closely i.e., by graphical representation showing several parameters of the
stars, such as the  radial and transverse pressures radius, energy density the gravitational mass,
compactness of star, and some other interesting quantities.

 Here, we observe that $\rho_c $, $ p_{rc}$ and $p_{tc}$
is maximum at the origin and it decreases radially outward, which is
the most obvious effect that needs to be considered in Fig. \textbf{2} .
On the other hand,  the energy density is continuous and well behaved in the stellar interior has
been shown in Fig. \textbf{3} whereas in Fig. \textbf{3} (right panel) shows the dependence of the anisotropic
nature. For a view on the status of the compactness and gravitational redshift function of strange compact stars shown in Fig. \textbf{4} exhibit several basic properties ordinary
anisotropic matter distribution.

The entire analysis has been performed with a set of astrophysical objects in
connection to direct comparison of some strange/compact star candidates
like X-ray pulsar PSR J1614-2230, PSR J1903+327 , X-ray burster 4U 1608-52, and Xray sources Vela X-1 .
A summary of numerical values of physical parameter for compact star candidates is given in Table I.

 \begin{figure}[h!] \centering
	\includegraphics[width=5.5cm]{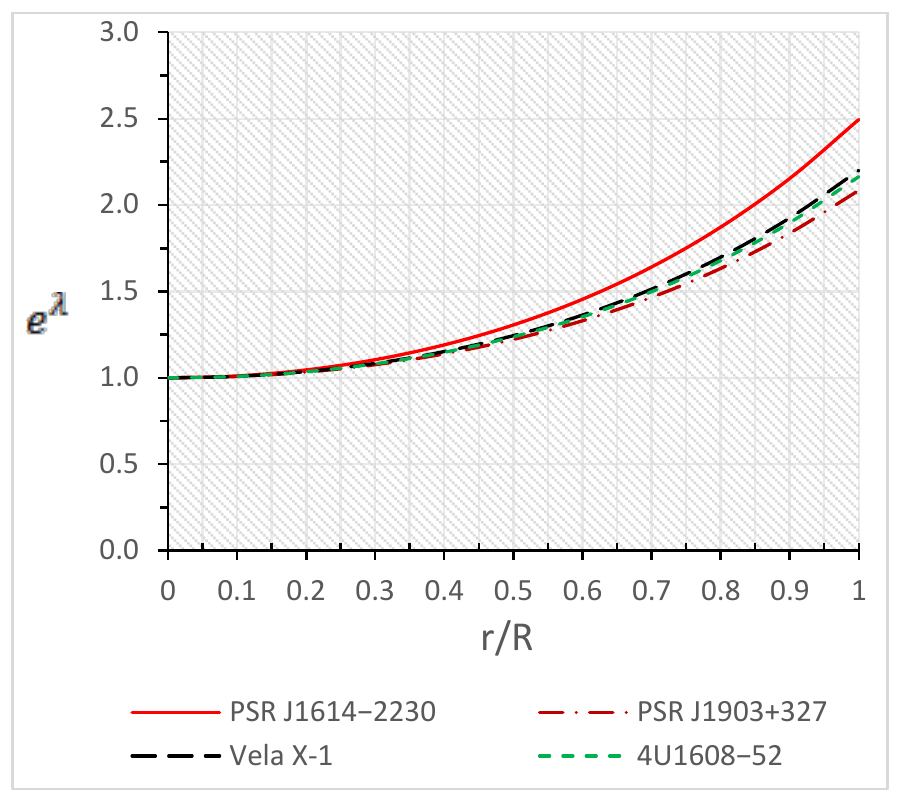} \includegraphics[width=5.5cm]{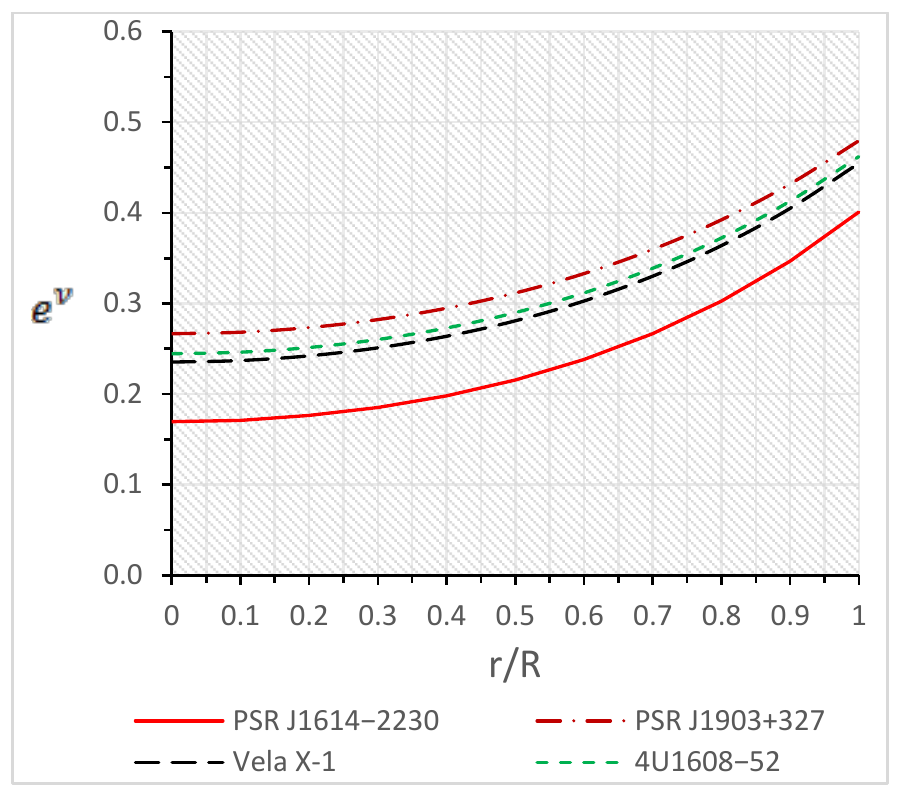}
	\caption{variation of matric functions $e^{\lambda}$ (left panel) and  $e^{\nu}$ (right panel) against $r/R$.}
	\label{11}
\end{figure}

\begin{figure}[h!] \centering
	\includegraphics[width=5.5cm]{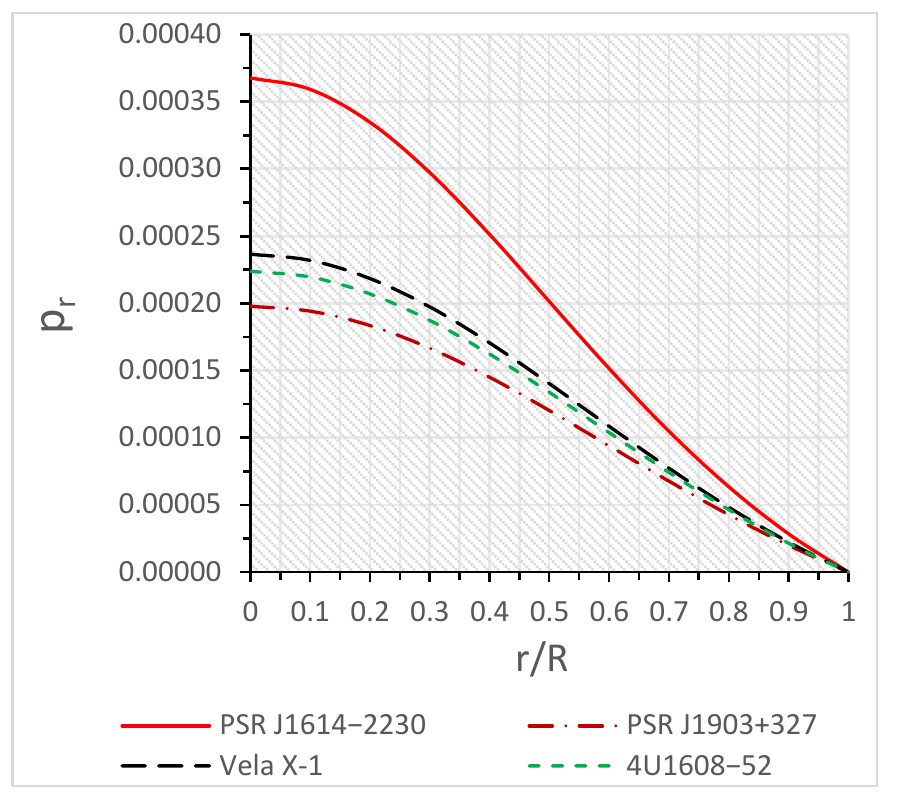} \includegraphics[width=5.5cm]{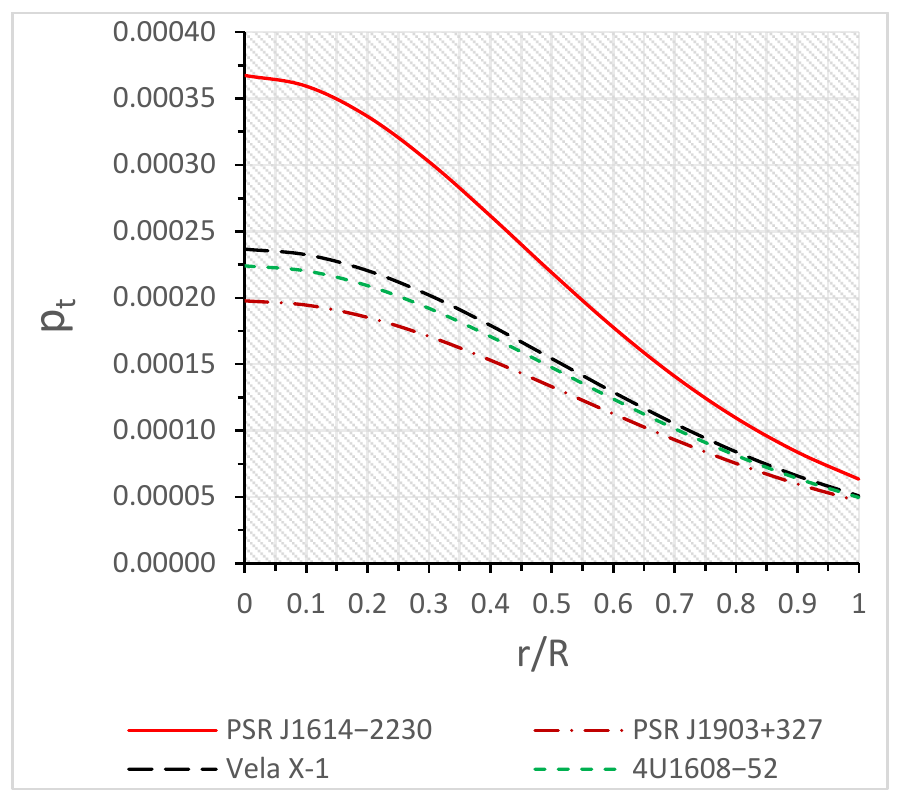}
	\caption{variation of radial pressure $p_r$ (left panel) and tangential pressure $p_t$ (right panel) against $r/R$.}
	\label{11}
\end{figure}

 \begin{figure}[h!] \centering
	\includegraphics[width=5.5cm]{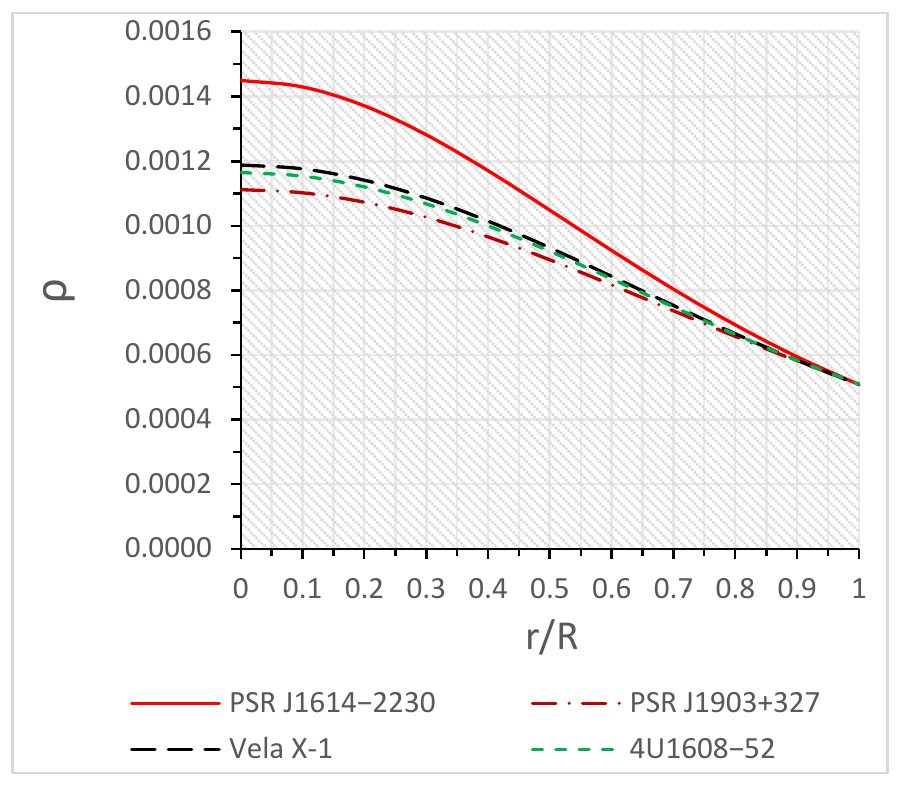} \includegraphics[width=5.5cm]{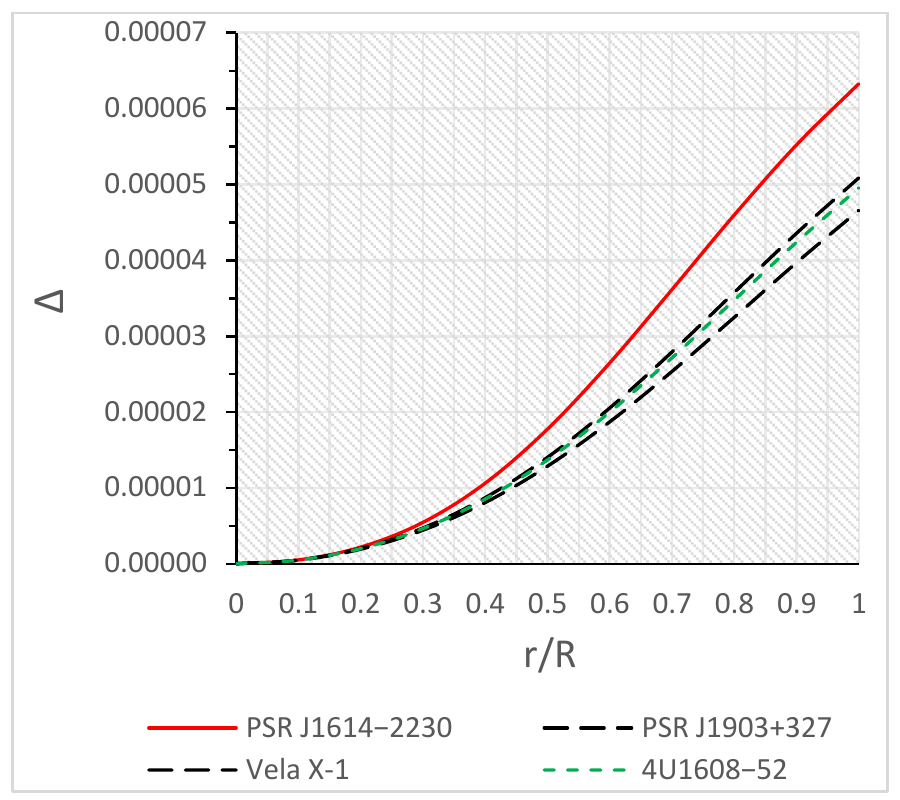}
	\caption{In the left panel the variation of density $\rho$ against $r/R$  (left panel),
whereas in the right panel variation of anisotropy factor $\Delta$ against $r/R$ has been considered.}
	\label{11}
\end{figure}
\begin{figure}[h!] \centering
	\includegraphics[width=5.5cm]{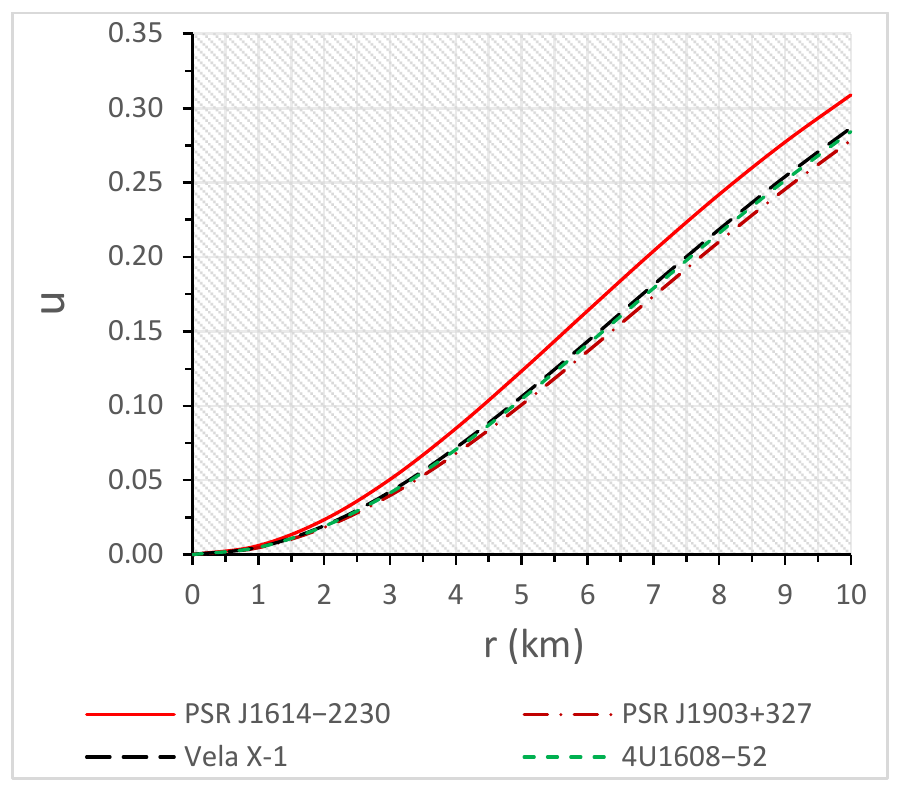} \includegraphics[width=5.5cm]{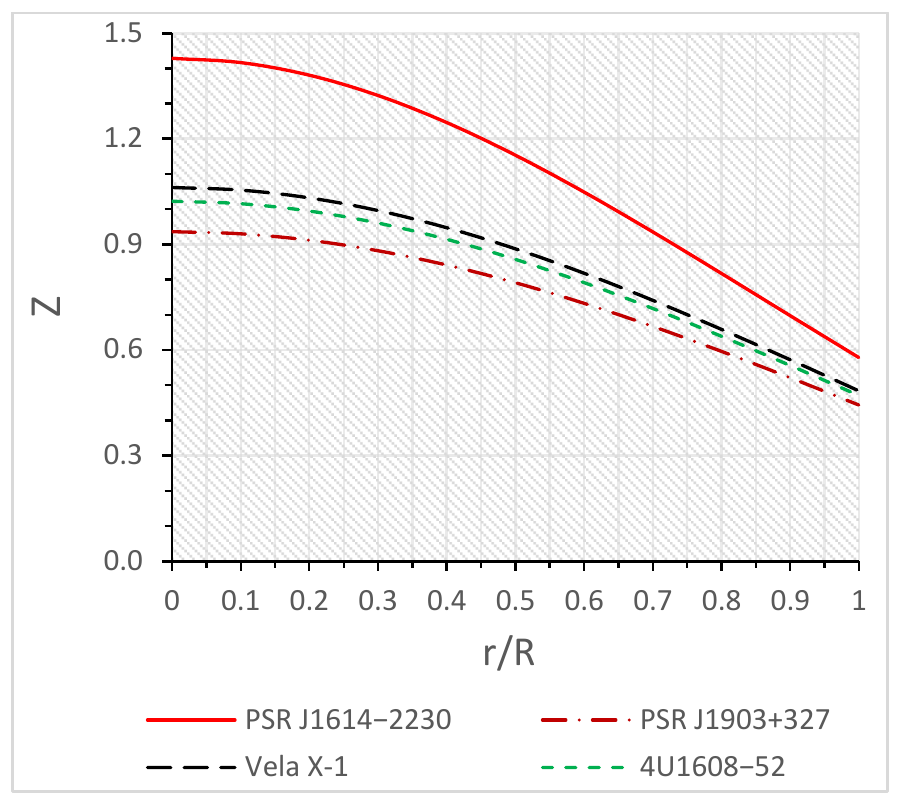}
	\caption{variation of the compactness and redshift against $r/R$ . }
	\label{11}
\end{figure}

\section{Physical properties and Comparative study of the physical parameters for
compact star model}
From an astrophysical point of view the general scenario presented above is extremely intriguing and
in this section we shall investigate how stellar solution must satisfy some general physical requirements.
Therefore our motivation  is to explore the physical features of the compact star and determine the constraints for which the solutions are physically realistic.

\subsection{Equilibrium under three different forces}
It is interesting to study the equilibrium conditions under the different forces, namely, gravitational, hydrostatic and anisotropic forces, respectively. So, there is only the need to consider the generalized Tolman-Oppenheimer-Volkov (TOV) equation for anisotropic fluid distribution, given by \cite{leon,Das}
\begin{equation}\label{p1}
-\frac{M_G(\rho+p_r)}{r^{2}}~e^{(\lambda-\nu)/2}-\frac{dp_r}{dr}+\frac{2(p_t-p_r)}{r}=0,
\end{equation}
where the effective gravitational mass $M_G(r)$  inside the fluid sphere of radius `$r$'  is given by :
\begin{equation}
M_G(r)=4\pi \int_0^r \Big(T_0^0-T_1^1-T_2^2-T_3^3 \Big)~r^2 e^{(\nu+\lambda)/2} dr, \label{wt}
\end{equation}
and, for notational convenience, the factors may be written as
\begin{equation}
M_G(r)=\frac{1}{2}r^{2}\nu'~e^{(\nu-\lambda)/2}.
\end{equation}
Let us now attempt to understand the Eq. (32) from an equilibrium point of view of a static spherically symmetric body and is usually done by proceeding from some reasonable forces, namely,  gravitational,
hydrostatic and anisotropic forces for stellar objects, which may be rewritten as follows:
\begin{equation}
F_g+F_h+F_a=0,
\end{equation}
where $F_g =-\frac{\nu'}{2}(\rho+p_r)$, $F_h =-\frac{dp_r}{dr}$ and $F_a=\frac{2}{r}(p_t-p_r)$
denotes the the gravitational, hydrostatic and anisotropic forces, respectively.
Theoretically a convenient way of expressing this equilibrium position by
graphical representation as evidenced in Fig. \textbf{5}.

 \begin{figure}[h!] \centering
	\includegraphics[width=5.5cm]{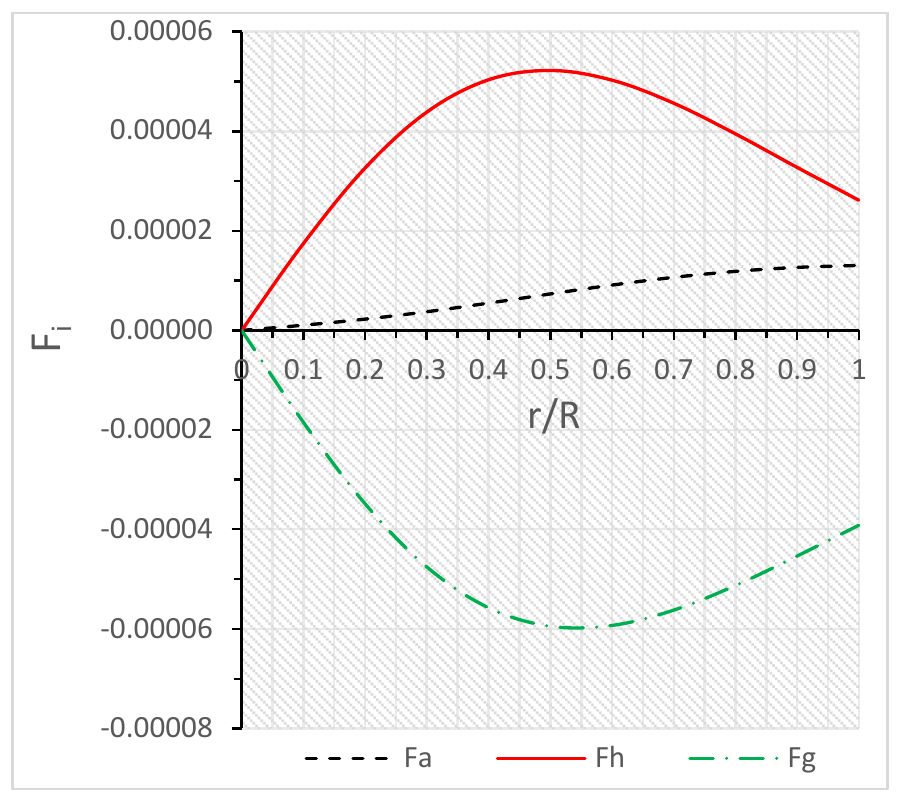} \includegraphics[width=5.5cm]{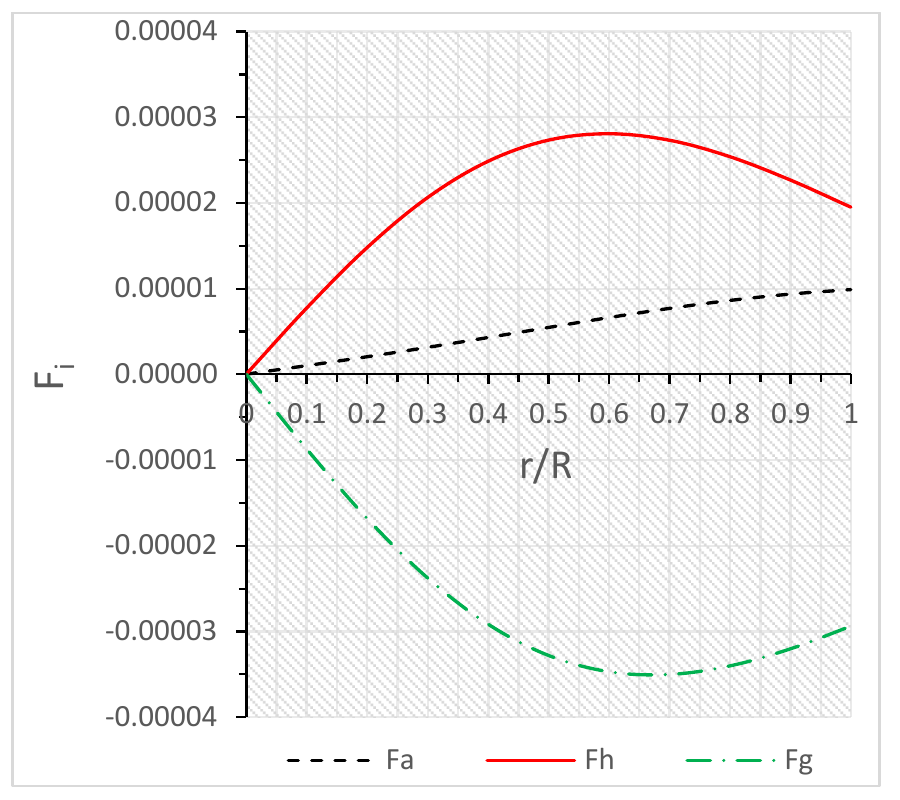}
    \includegraphics[width=5.5cm]{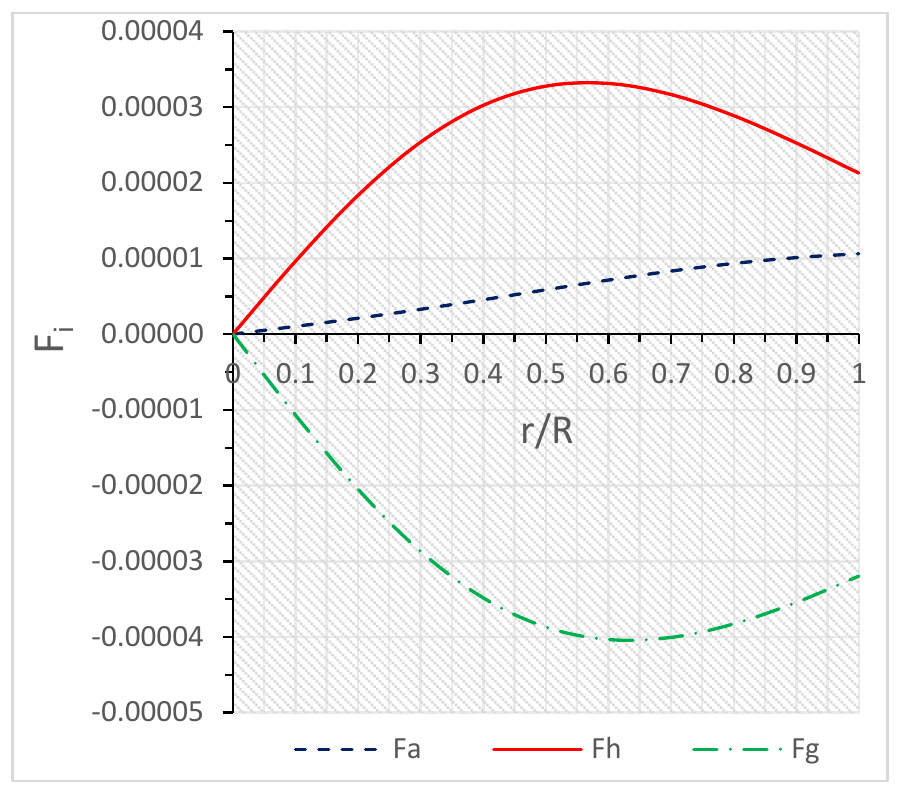} \includegraphics[width=5.5cm]{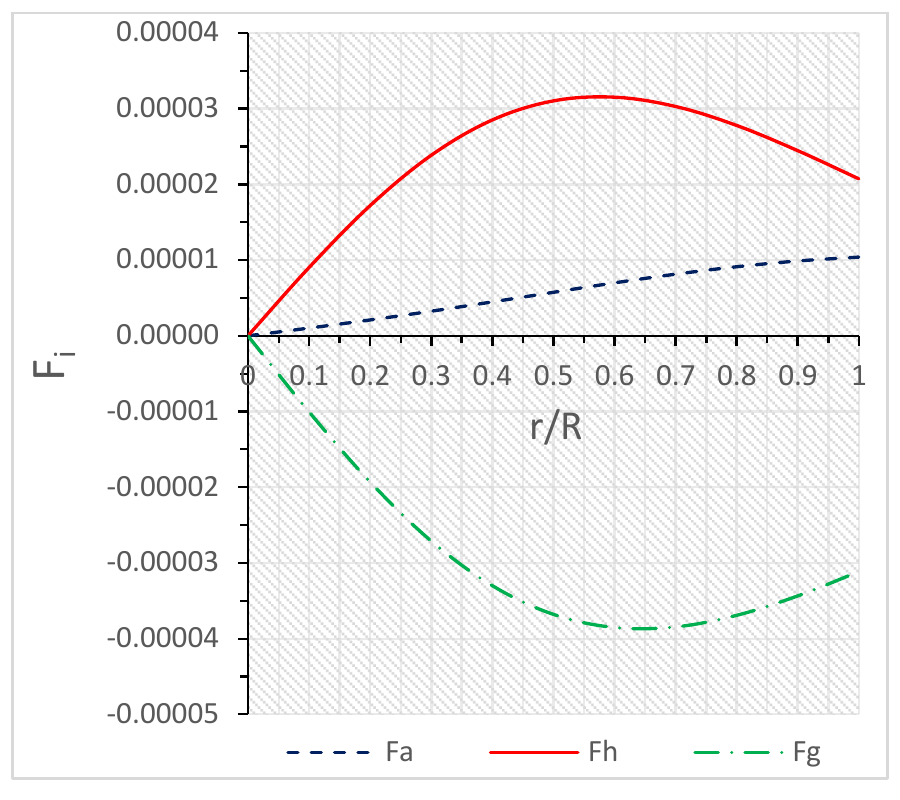}
	\caption{variation of different forces $F_i$  against $r/R$.}
	\label{11}
\end{figure}

\subsection{Causality condition and stability}
Here we are interested in checking the velocity of sound, using the concept of Herrera's
cracking (or overtuning) \cite{Herrera}. According to the
causality condition implies that the radial and tangential speed of sound should be less than unity
i.e., $0 < v_{r}^{2}\leq 1$ and $0 < v_{t}^{2}\leq 1$,  everywhere within the stellar object where
\begin{equation}
v_{r}^2=\frac{dp_r}{d\rho}={dp_r/dr \over d\rho/dr},~~~v_{t} ^2=\frac{dp_t}{d\rho}={dp_t/dr \over d\rho/dr}.
\end{equation}
We consider the graphical representation ( see Fig. [6-7]) for more closely look into
the radial and transverse velocity of sound that obeys the causality conditions, i.e., both $v_r^2,\,v_t^2$ are less than unity and monotonic decreasing function.

A consideration is in order at this point to be mentioned that the  anisotropic fluid sphere with $-1\leq v_{t}^{2}-v_{r}^{2}\leq 0$ is potentially stable, whereas the region $0<v_{t}^{2}-v_{r}^{2}\leq 1$ is potentially unstable. In this cases the sound speed which is $v_t^2-v_r^2<0$ is finite
everywhere inside the fluid sphere (See Fig. 8).

\begin{figure}[h!] \centering
	\includegraphics[width=5.5cm]{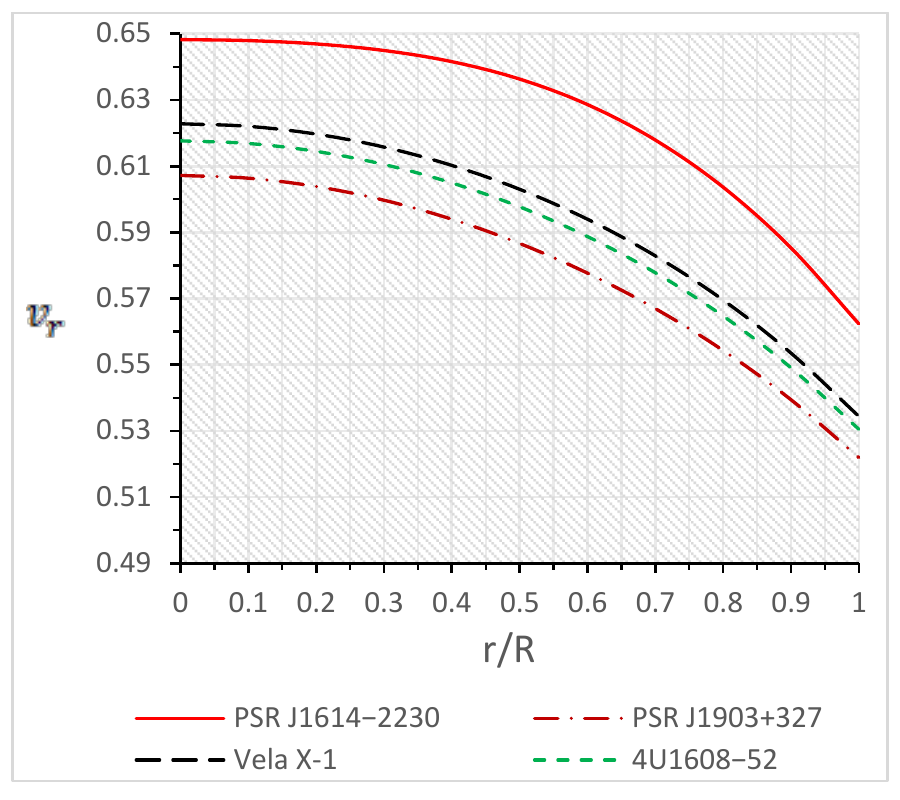} \includegraphics[width=5.5cm]{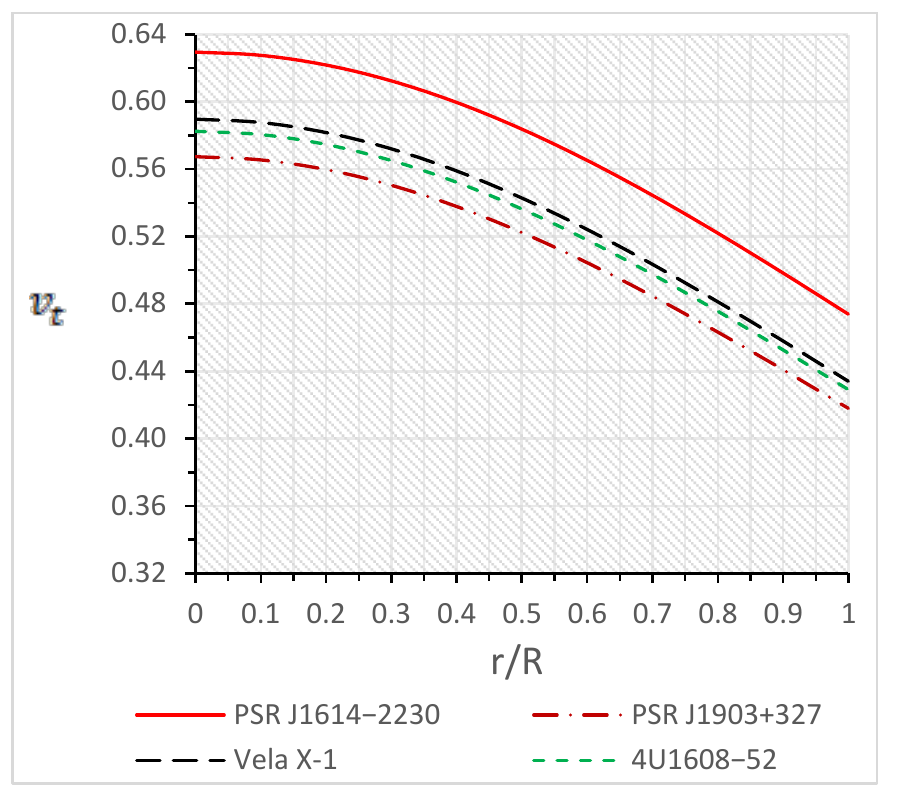}
	\caption{variation of radial velocity $v_r$ (left panel) and tangential velocity $v_t$ (right panel) against $r/R$.}
	\label{11}
\end{figure}

\begin{figure}[h!] \centering
	\includegraphics[width=5.5cm]{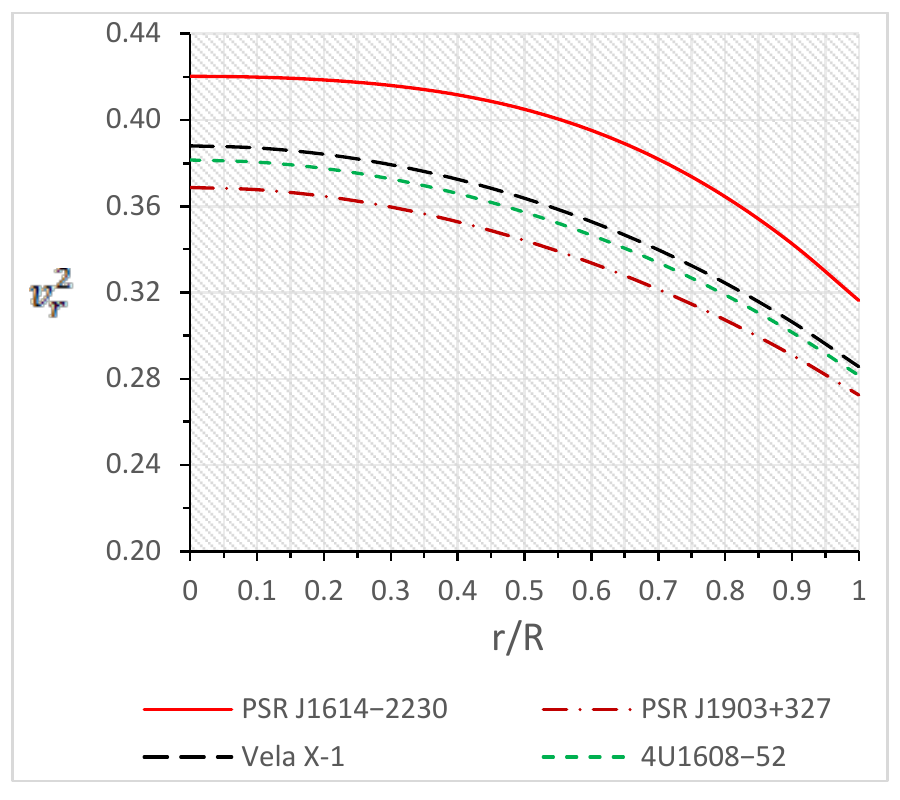} \includegraphics[width=5.5cm]{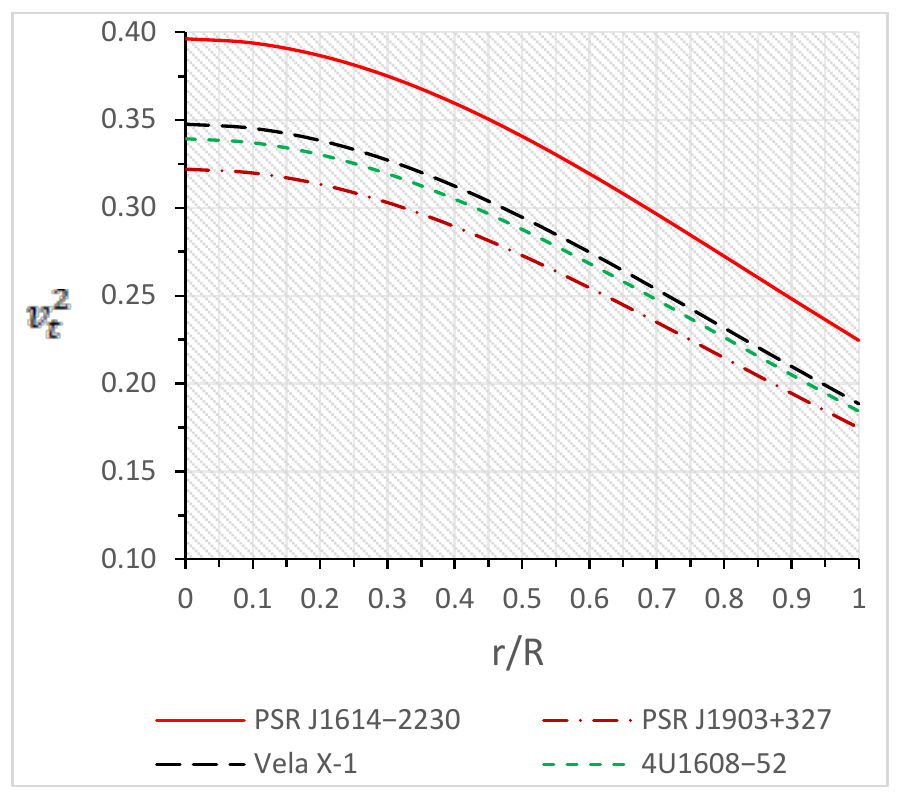}
	\caption{variation of square radial velocity $v^2_r$ (left panel) and tangential velocity $v^2_t$ (right panel) against $r/R$.}
	\label{11}
\end{figure}

\begin{figure}[h!] \centering
	\includegraphics[width=5.5cm]{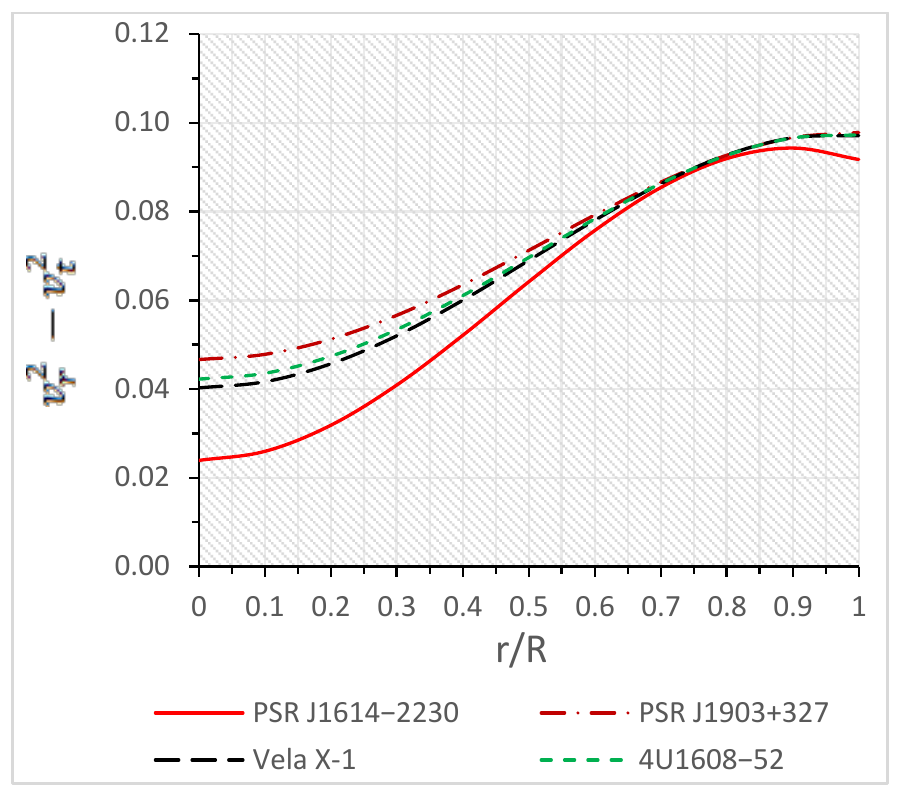} \includegraphics[width=5.5cm]{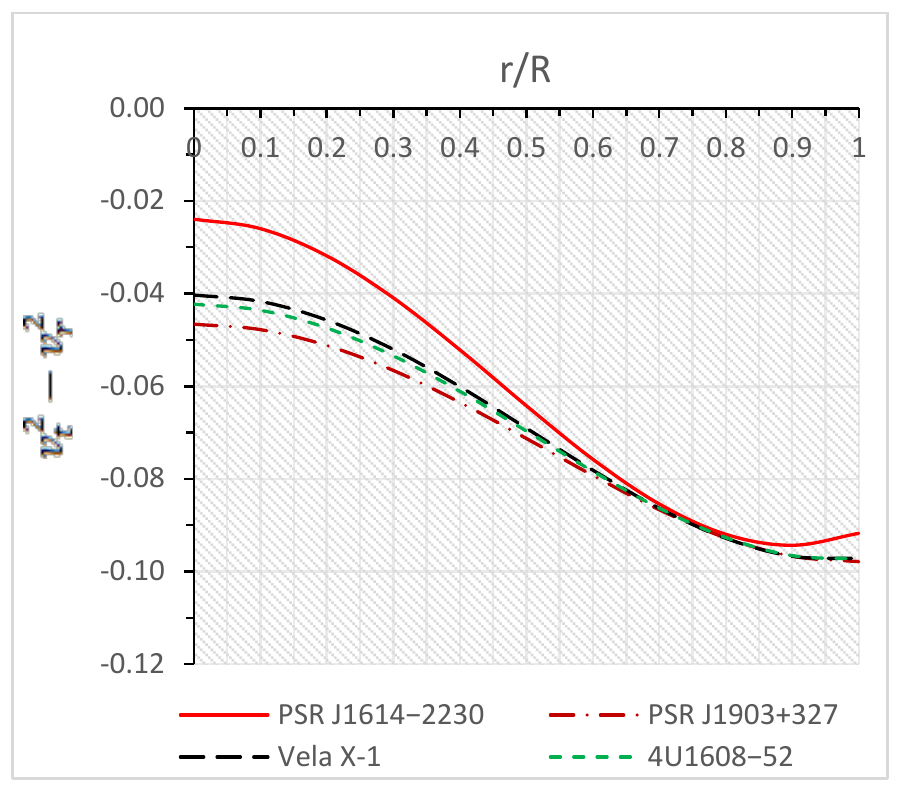}
	\caption{variation of difference of velocities  $v^2_r-v^2_t$ (left panel) and $v^2_t-v^2_r$ (right panel) against $r/R$.}
	\label{11}
\end{figure}

 \subsection{Relativistic adiabatic index and Stability}
Due to the anisotropic matter distribution of compact
astrophysical objects the expression for adiabatic index $\Gamma$, is given by \cite{Chan}
\begin{equation}
\Gamma=\frac{\rho+p_r}{p_r}\frac{dp_r}{d\rho},
\end{equation}
which can relate the stability of the anisotropic configuration.
In the case of stable relativistic star, the adiabatic index must be $\Gamma$ $>$ 4/3
throughout the configuration. In the case of a anisotropic fluid sphere
the adiabatic index is equivalent to \\

$\Gamma=-\left[1+\frac{4bB-a\left(A+B\ln\cosh\psi\right)\tanh\psi}{a\left[A+B\ln\cosh\psi\right]\left[4\,b\,r^2\,sech^2\psi
  +3\tanh\psi+a\,r^2\tanh^3\psi\right]}\right] \times\\ \left[\frac{\left[f_{1}\left(\psi\right)f_{2}\left(\psi\right)+
  \tanh^2\psi\left(f_{3}\left(\psi\right)+a\,f_{4}\left(\psi\right)\right)\right]\left[1+a\,r^2\tanh^2\psi\right]}
  {\left[A+B\,\ln\cosh\psi\right]^2\left[f_{13}\left(\psi\right)+f_{14}\left(\psi\right)\right]}\right].$\\

The above-mentioned condition would be changed for a relativistic isotropic sphere
due to the regenerative effect of pressure, which renders the sphere more unstable.
Generally For an anisotropic fluid distribution situation becomes more complicated, because the stability
will depend different types of  pressures.

\begin{figure}[h!] \centering
	\includegraphics[width=5.5cm]{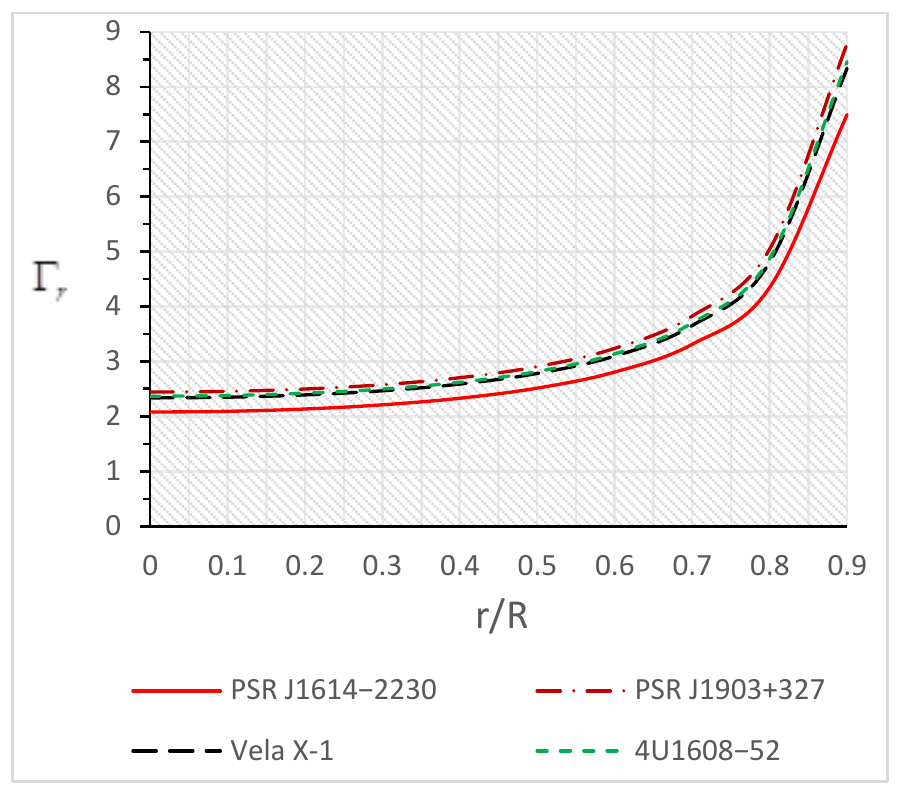}
	\caption{The relativistic adiabatic indices are potted against $r/R$.}
	\label{11}
\end{figure}

\subsection{\emph{Energy Condition}}
The general scenario presented above is extremely intriguing from an astrophysical point of view as our next goal is to verify whether our  static stellar configurations, satisfies all the energy conditions or not, namely, null energy condition (NEC), weak energy condition (WEC), strong energy condition(SEC) and dominant energy condition (DEC), at all points in the interior of a star. The next step is to write the
 inequalities in the following form:
\begin{eqnarray}
\textbf{NEC:}~ \rho(r)-p_r \geq  0,\\
\textbf{WEC:}~ \rho(r)-p_r(r) \geq  0~~, ~~\rho \geq  0,\\
\textbf{SEC:}~ \rho(r)-p_r(r) \geq  0~~, ~~\rho-p_r(r)-2p_t(r) \geq  0.
\end{eqnarray}
For this purpose, we plot four diagrams related to different energy conditions
versus radius for different compact star candidates in Figs. \textbf{9}. Note that we
only write down the inequalities and plotted the graphs because of complexity of the expression.

 \begin{figure}[h!] \centering
	\includegraphics[width=5.5cm]{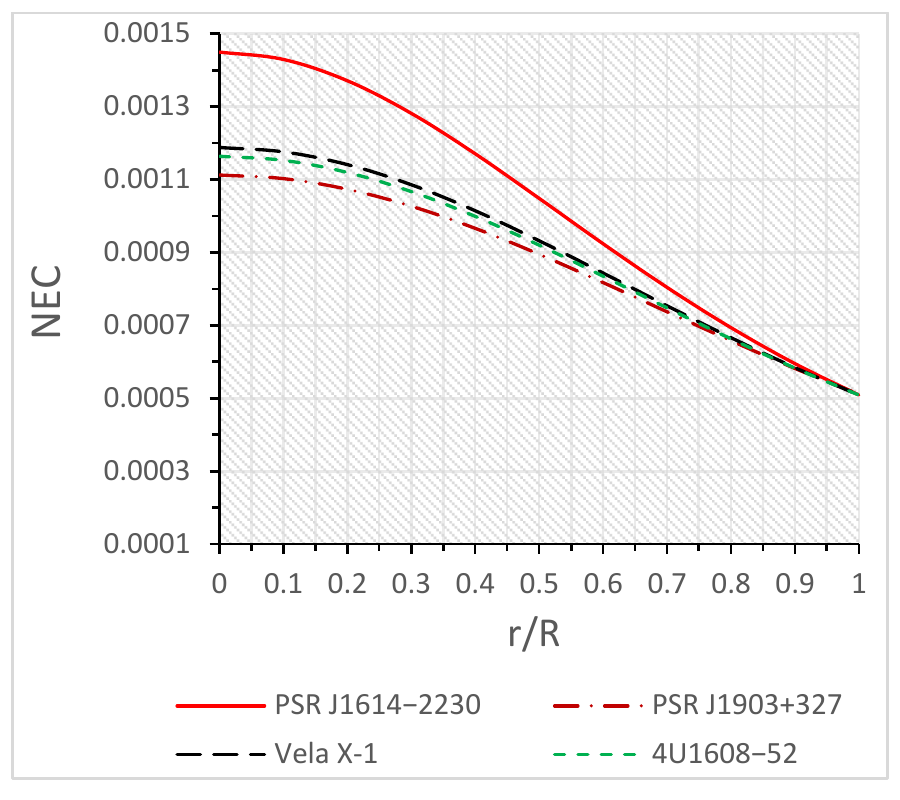} \includegraphics[width=5.5cm]{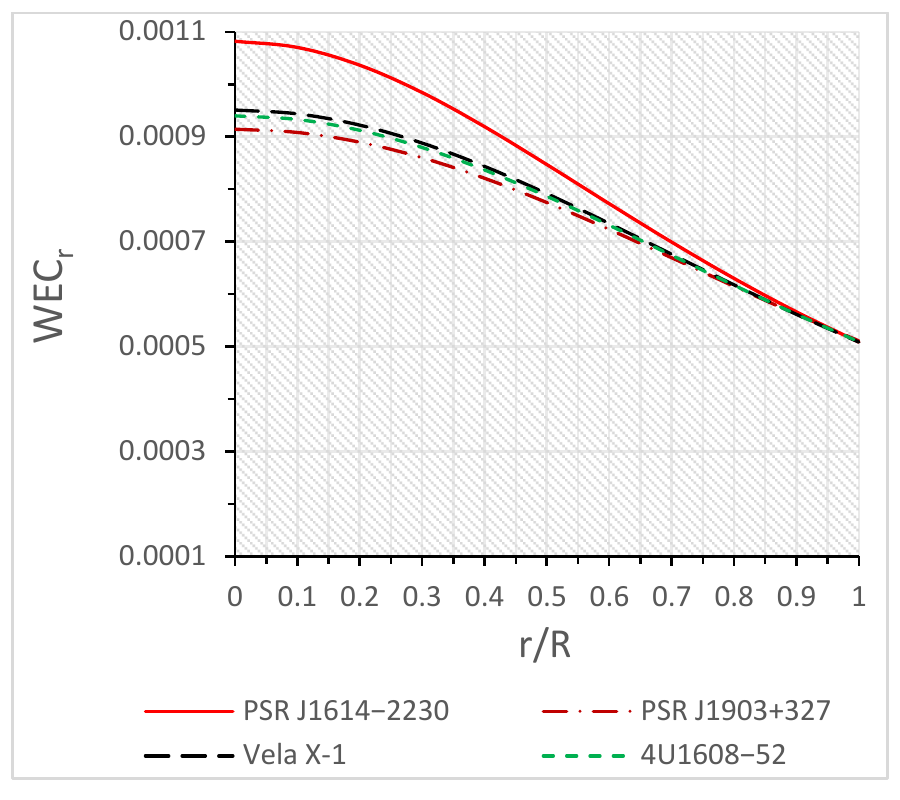}
    \includegraphics[width=5.5cm]{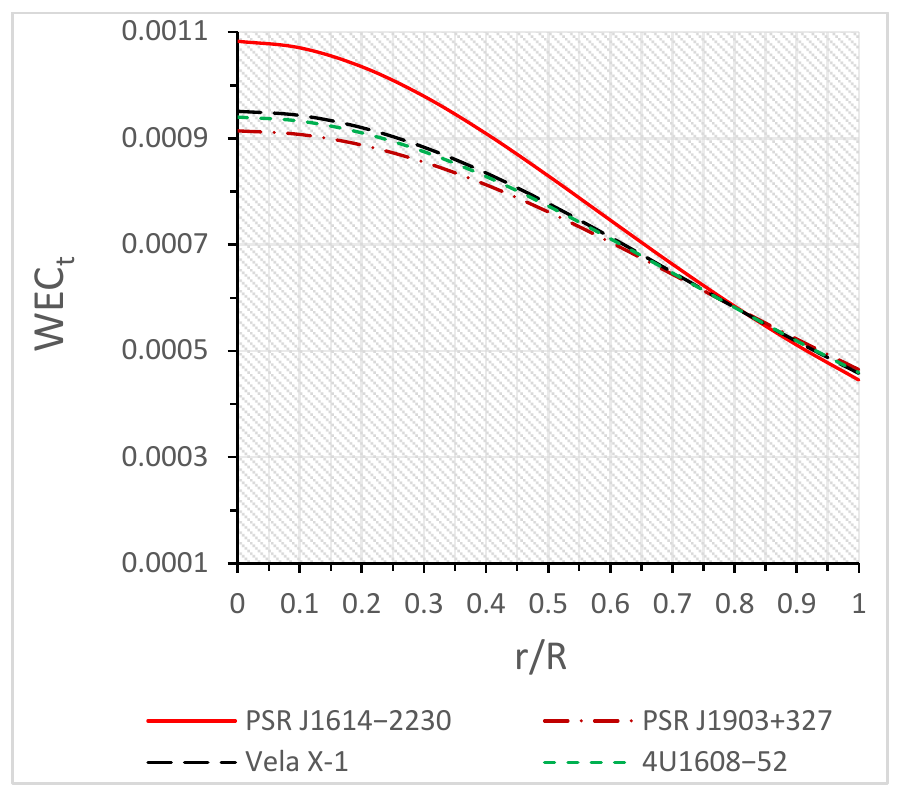} \includegraphics[width=5.5cm]{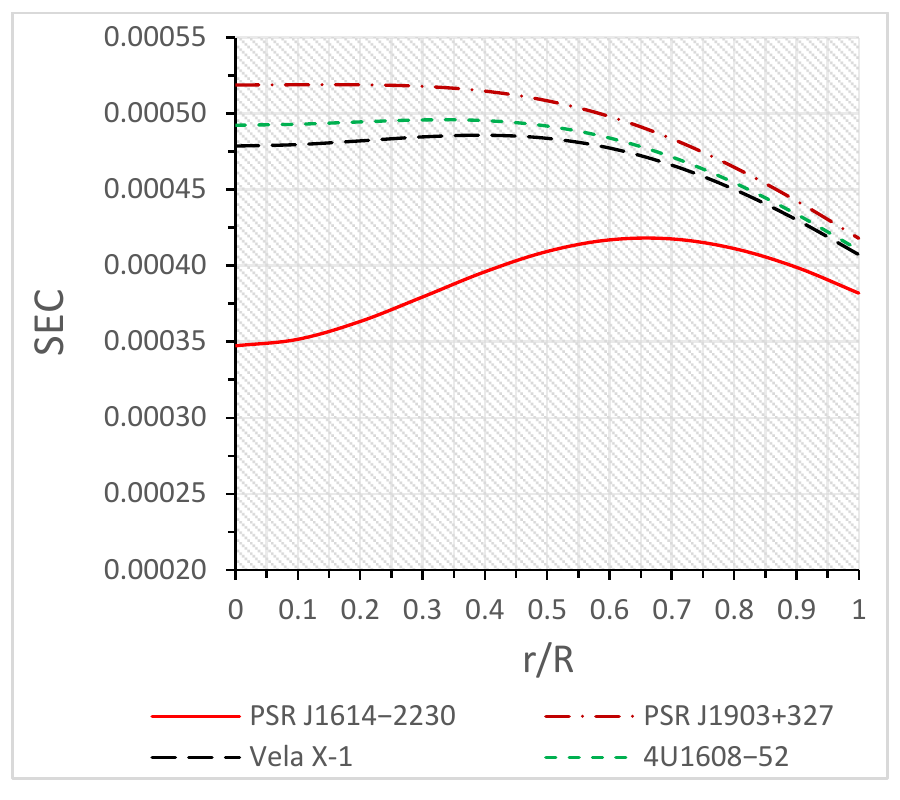}
	\caption{variation of energy conditions against $r/R$.}
	\label{11}
\end{figure}

\begin{table}

\centering
	\caption{Values of the model parameters Mass Radius($R$), ($M_{\odot}$), $a$ $b$, $c$, $A$ $B$  for different compact stars}

\resizebox{\columnwidth}{!}{\begin{tabular}{@{}lrrrrrrrr@{}} \hline
Compact Star & R(km) & M$\left(M_\odot\right)$ & $a\,(km^{-2})$ & $b\,(km^{-2})$ & $c$& A & B\\
\hline PSR J1614-2230  & 9.69 & 1.97 & 0.2911 & 0.00033 & 0.2072 & - 0.2842097 & 32.6469207\\
\hline PSR J1903+327 & 9.438 & 1.667 & 0.2727 & 0.00031 & 0.1870& - 0.043234 & 32.191057\\
\hline Vela X-1 & 9.56 & 1.77 & 0.2803 & 0.00032 & 0.1907& - 0.0926653 & 31.95616\\
\hline 4U1608-52 & 9.528 & 1.74 & 0.2804 & 0.000315 & 0.1887& - 0.07835 & 32.35636\\
\hline
\end{tabular}}
\end{table}

\begin{center}
\begin{tabular}{ @{}l|rrrrr@{}  }
 \hline Table~2 & \multicolumn{4}{c}{Estimated physical values based on the observational data}\\ \hline
Compact star & $\rho_{0} (gm/cm^{3})$ & $\rho_{R} (gm/cm^{3})$ & $p_{c} (dyne/cm^{2})$ &  $M/R$\\ \hline
\hline PSR J1614-2230 & 1.957087$\times{{10}^{15}}$ & $6.861709\times{{10}^{14}}$ & $4.46592\times{{10}^{35}}$ & 0.3558 \\

\hline PSR J1903+327 & $1.5011397\times{{10}^{15}}$ & $6.89797\times{{10}^{14}}$ & $2.403309\times{{10}^{35}}$ & 0.3091 \\
\hline Vela X-1  & $1.603168\times{{10}^{15}}$ & $6.86923\times{{10}^{14}}$ & $2.87261\times{{10}^{35}}$ & 0.3240 \\
\hline 4U1608-52  & $ 1.571059\times{{10}^{15}}$ & $ 6.873648\times{{10}^{14}}$ & $ 2.720805\times{{10}^{34}}$ & 0.3196\\
\hline
\end{tabular}
\end{center}

\section{Concluding remarks}

In this work, we proposed a new model of compact star within embedding class one spacetime. The study has been performed assuming a new type of mass function. The reason to adopt such an approach is to investigate a singularity free  solutions i.e. solutions are well behaved at the centre of the star. However, we not only restricted out attention for maintaining only the physical criteria but also we emphasize the results in more details, of the existence of compact star as recently pointed out by observations \cite{gan}.

For clarity, we  explore several aspects of the model analytically along with graphical display in order to verify a model can be considered viable within the specified observational constraint that can describe a large class of compact stars. To place our newly developed star model we emphasize the results in more details, with observational data taken by Gangopadhyay {\em et al.} \cite{gan}. For that we consider observed masses of the compact objects, namely, PSR  J1614-2230, PSR  J1903+327,  Vela X - 1 and  4U 1680-52.

By using observational data sets for the radii and mass-radius relations,
we carry out a comparative study between the data of the model parameters with that of the compact star
candidates in Table~I and ~II. We choose $G = c = 1$,  while solving Einstein's
equations as well as for plotting all the figures. An important feature of the
present analysis is the obtained results are very much compatible with
the results obtained through the observations and the mass-radius
for different strange stars are lies within the proposed range by Buchdahl \cite{Buchdahl}.
Then we turned our attention to the equilibrium and stability conditions of stars, like
under different forces, adiabatic index and by checking the speeds of sound. In fact,
it is interesting to note that our present model successfully cross all the barriers.
These effects can lead to a number of distinctive compact objects, however, it is important
to understand the nature and general properties of compact stars more concisely.

\section*{Acknowledgments} The author S. K. Maurya acknowledges authority of University of Nizwa for their continuous support and encouragement to carry out this research work. FR and AB are thankful to the authority of Inter-University Centre for Astronomy and Astrophysics, Pune, India for providing research facilities.  FR is also grateful to DST-SERB and DST-PURSE,  Govt. of India for financial support. MR is also thankful to CSIR, Govt. of India for financial support.\\

\end{document}